# Possible depth-resolved reconstruction of shear moduli in the cornea following collagen crosslinking (CXL) with optical coherence tomography and elastography.

Gabriel Regnault, Mitchell A. Kirby, Ruikang K. Wang, Tueng T. Shen, Matthew O'Donnell, and Ivan Pelivanov.

**Abstract—** **Collagen crosslinking of the cornea (CXL) is commonly employed to prevent or treat keratoconus. Although the change of corneal stiffness induced by CXL surgery can be monitored with non-contact dynamic Optical Coherence Elastography (OCE) by tracking mechanical wave propagation, the depth dependence of this change is still unclear if the cornea is not crosslinked through the whole depth. Here we propose to combine phase-decorrelation measurement applied to OCT structural images and acoustic micro-tapping (AµT) OCE to explore possible depth reconstruction of stiffness within crosslinked corneas in an *ex vivo* human cornea sample. The analysis of experimental OCT images is used to define the penetration depth of CXL into the cornea, which varies from ~100µm in the periphery to ~150µm in the central area and exhibits a sharp transition between areas. This information was used in a two-layer analytical model to quantify the stiffness of the treated layer. We also discuss how the elastic moduli of partially CXL-treated cornea layers reconstructed from OCE measurements reflect the effective mechanical stiffness of the entire cornea to properly quantify surgical outcome.**

## I. INTRODUCTION

AT the interface between air and the inner eye, the cornea provides protection and is the primary optical element focusing light onto the retina. It contains multiple layers, including epithelium and stroma. The first acts as a barrier against the external environment and the latter maintains its stiffness, transparency and focusing power [1], [2]. The microstructure of the stroma is composed of collagen fibrils, arranged in lamellae, lying within a protein rich, hydrated proteoglycan mesh [3], [4].

Corneal diseases (such as keratoconus (KC)) and surgical complications from refractive surgeries (such as LASIK) may deform the cornea (ectasia) and reduce vision. The prevalence of KC in the general population is estimated to be 1.38 per 1000 [5], and nearly 1 million refractive surgeries are performed each year in the USA. Despite their overall success, however, suboptimal visual outcomes and post-refractive corneal decompensation cannot always be predicted for an individual patient. Corneal collagen crosslinking (CXL) is a minimally invasive procedure that can potentially slow the progression of corneal ectasia [6]–[9]. UV light modifies the microstructure of the cornea soaked in riboflavin and forms additional chemical bonds between collagen fibers in the stroma [10]. Post-treatment corneas become stiffer and more resistant to enzymatic digestion [11]–[13]. Although corneal topography (curvature map) and thickness map can be obtained preoperatively, and needed refractive corrections can be estimated, there is an unmet need to predict corneal decompensation from interventions such as LASIK and CXL therapies. Unfortunately, surgical planning cannot be customized and outcomes (e.g. post-surgery corneal ectasia risks) cannot be accurately predicted without quantitatively mapping corneal elasticity. Thus, methods that can quantitatively reconstruct corneal elastic moduli are in high demand.

Ocular response analyzer (ORA - Reichert Technologies) and Dynamic Scheimpflug Analyzer (DSA - Corvis ST – Oculus Opitkgerate GmbH) are the state-of-the-art in clinical measurements of corneal mechanics. They estimate stiffness as the pressure at inward applanation divided by corneal displacement [14]–[16]. Over 100 papers are published annually on the topic, and the Journal of Cataract and Refractive Surgery devoted an entire issue to the topic [15]. However, measurements induce large corneal deformations that are often clinically unacceptable, require a non-trivial IOP correction in simulations [17] and assume a simple isotropic mechanical model leading to high variability with experimental conditions. Results obtained with the Corvis ST on KC may be contradictory, and some even show no significant change in corneal stiffness pre- and post-CXL surgery [18], [19]. In addition, the result is averaged over the entire cornea with no spatial resolution, and the reconstruction is questionable if corneal thickness varies.

There is no consensus in the literature on the elastic model for cornea and the stiffness range even for healthy subjects. The most common model assumes an incompressible, isotropic, and linear elastic solid, where a single parameter, the Young's modulus $E$ (or equivalently the shear modulus $\mu = E/3$), defines elasticity. Destructive mechanical tests can determine E *ex vivo*, with reported values for human cornea (at low-strain) of 800 kPa to 4.7 MPa for tensile loading [12], [20]–[23], and 100 kPa to 3 MPa for inflation loading [24], [25]. Note that the destructive nature of mechanical tests precludes their clinical translation.

Dynamic elastography is a promising tool to probe soft tissue biomechanics. A shear wave can be launched using direct contact excitation [26]–[29] or radiation force-based techniques [30], [31]. By tracking shear wave propagation, either using MRI [32], [33] ultrasound [26]–[28], [34] or dynamic phase-sensitive OCT techniques [35]–[37], one can infer, with an appropriate mechanical description, the linear [38]–[41], or non-linear [42], [43], stiffness moduli of the tissue. Optical coherence elastography (OCE) is particularly suited to probe corneal biomechanics non-invasively in a clinical environment [30], [36], [44]–[46], as it can be combined with non-contact excitation techniques (for example, using an air-puff or acoustic micro-tapping (AμT [30])).

Because cornea is thin and bounded, above by air and below by aqueous humor, wave propagation within it is guided, leading to strong geometric dispersion [39], [47]. As such, the common approach associating the Rayleigh surface wave group velocity to stiffness [30], [36], [44]–[46], [48]–[51] is not appropriate. It in fact results in a 2 order of magnitude mismatch in Young's modulus compared to tensile and inflation tests [47]. Accounting for phase-velocity dispersion is mandatory to recover corneal stiffness. Using Lamb-wave dispersion, OCE studies reported corneal shear moduli in the range of $1.8 - 52.3$ kPa [49], [52], [53], in close agreement with values obtained from torsional and rheometry testing of *ex vivo* cornea ($2.5 - 47.3$ kPa) [54]–[56]. However, all shear-based methods produce moduli differing by 1-2 orders of magnitude from those reported by tensile and inflation tests.

Recently, we hypothesized that corneal anisotropy is the primary cause of these discrepancies [39]. Corneal microstructure supports this hypothesis. The stroma contains collagen lamellae running in-plane across its width. Lamellae make up approximately 90% of tissue thickness and account for most of the cornea's mechanical structure. They are stacked vertically in approximately 200-500 separate planes [57], [58], suggesting an anisotropic mechanical behavior with very different responses to in-plane versus out-of-plane loads. We introduced a model of a nearly-incompressible transverse isotropic (NITI) medium [39], in which corneal stiffness is defined by two (in-, $\mu$, and out-of-plane, $G$) shear moduli, decoupling tensile/inflation properties from shear responses. Based on this model, we developed an algorithm utilizing guided mechanical waves in a bounded NITI medium to reconstruct both moduli from AμT-OCE. The model was confirmed *ex vivo* in rabbit [59], porcine [60] and human [38], [39] models and *in vivo* with rabbit models [59].

In [38] we showed that both in- and out-of-plane post-CXL corneal shear moduli increased compared to their pre-surgery values, with an averaged two-fold increase in Young's modulus and an almost four-fold increase for the out-of-plane shear modulus $G$. This confirmed that CXL better crosslinks corneal lamellae. However, deformations under physiological conditions are defined by the Young's modulus, which is less affected and has implications for potential refractive changes. Note, however, that the effects of crosslinking are always non-homogeneous in depth, as riboflavin penetration is more pronounced in the anterior region of the cornea where the solution is applied during soaking [61], and because UVA irradiation is more efficient on the anterior part of the cornea [62]. Together, this generally leads to a clear demarcation line, indicating the efficiency of the treatment [63]. Note that riboflavin penetration is purposely limited to minimize damage to the endothelium and ensure the biological integrity of the cornea [6].

As noted above, mechanical waves generated in the cornea are guided. Such guided waves occupy the entire depth of the cornea and, therefore, carry average information over depth. Thus, reconstructing the depth dependence of corneal moduli is practically impossible without a good estimate of the penetration of CXL into the cornea. Blackburn et al. [64] have recently introduced a novel metric to track CXL penetration within the cornea using time-resolved OCT. They demonstrated that the phase decorrelation decay rate of the complex OCT signal is reduced in CXL areas and can be used to distinguish treated and untreated areas postoperatively.

In this paper, we combine the methods described in [64] with AμT-OCE measurements to explore possible reconstruction of both in- and out-of-plane corneal elastic moduli over depth. For this purpose, we developed an analytical model of guided wave propagation accounting for multiple layers in the cornea, each with distinct stiffness moduli and thickness. In addition, elastic moduli prior to CXL can be directly estimated from AμT-OCE so that procedure induced changes can be quantified. We also discuss how the elastic moduli of a CXL-treated layer reconstructed from OCE measurements influences the effective mechanical stiffness of the entire cornea to predict its deformational properties and properly quantify surgical outcome.

## II. METHOD

### A. Cornea preparation

One corneal-scleral ring, stored in Optisol (Chiron Ophtalmics) was obtained from CorneaGen. This sample from a 26 years-old donor was stored for less than 30 days. The corneal-scleral button is mounted on an artificial anterior chamber (Barron, CorzaMedical; see Fig. 1), connected through the inlet port to an elevated bath filled with balanced saline solution (BSS) to apply a controlled pressure on the anterior segment of the cornea. The outlet port remained closed to allow the pressure to settle at 15 mmHg within the chamber, corresponding to human physiological conditions [65]. Crosslinking was performed following the Dresden protocol [6]. First, the epithelial membrane was removed from the sample. Then, the cornea was soaked in riboflavin for 30 min by applying a 50μL drop of 0.1% riboflavin in 20% dextran solution every two minutes. The cornea was then exposed to 3mW/cm² of 370 nm ultra-violet (UV) light for 30 minutes, while a drop was re-applied every 5 minutes.

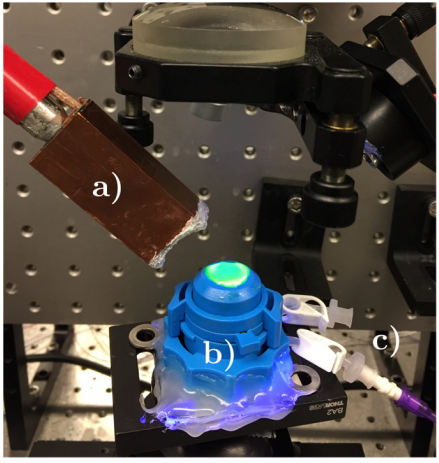

Figure 1. Picture of the experimental set up during UV-CXL. a) Acoustic micro-tapping transducer. b) Artificial anterior chamber with c) inlet port connected to the elevated bath and outlet port closed for controlling IOP.

### B. AµT-OCE imaging system

A spectral domain OCT system with a 46.5 kHz frame rate was used to track wave propagation and structural changes within the cornea. As described in previous studies [30], [36], [38], [39], a cylindrically focused air-coupled transducer, operating at a 1 MHz frequency, generated a spatio-temporal sharp displacement at the surface of the cornea using reflection-based acoustic radiation force. Because of the transducer's cylindrical geometry, the push was line-shaped and generated quasi-planar guided waves within bounded tissue. The OCT system operated in M-B mode. A single push was triggered by the system while 512 consecutive A-scans were taken at a fixed location (M-scan). The M-scan sequence and push excitation were repeated for 256 locations, creating a three-dimensional volume with 256 $x$-samples, 1024 $z$-samples and 512 $t$-samples (see Fig. 2a), with an effective imaging range of 6 mm × 1.2 mm × 11 ms. The vertical particle velocity was obtained from the optical phase difference between two consecutive A-lines at each location [66]. The spatio-temporal ($x$-$t$) surface signature of the guided wave was computed from an exponentially weighted-average of the vertical particle velocity over the first 180 µm of the anterior part of the cornea. As shown in Fig. 2b), the guided wave only propagated during the first 4 ms of the scans, which was used to determine the stiffness of the material by fitting the computed dispersion curve in the frequency-wavenumber domain ($f$-$k$) obtained from the 2D Fourier spectrum (Fig. 2c). This procedure is detailed in Section IID. On the other hand, data from the last 7 ms were used to study structural changes with phase decorrelation (see Section IIF).

### C. Multi-layer NITI model

Like most biological tissue, the cornea is nearly incompressible. In addition, its microstructure implies its transverse isotropy [60] and, therefore, its mechanical behavior under small deformation can be described with the NITI model [39]. In Voigt notation, Hook's law of stress and strain for a NITI material takes the form:

$$\begin{bmatrix} \sigma_{xx} \\ \sigma_{yy} \\ \sigma_{zz} \\ \tau_{yz} \\ \tau_{xz} \\ \tau_{xy} \end{bmatrix} = \begin{bmatrix} \lambda + 2\mu & \lambda & \lambda & & & \\ \lambda & \lambda + 2\mu & \lambda & & & \\ \lambda & \lambda & \lambda + \delta & & & \\ & & & G & & \\ & & & & G & \\ & & & & & \mu \end{bmatrix} \begin{bmatrix} \epsilon_{xx} \\ \epsilon_{yy} \\ \epsilon_{zz} \\ \gamma_{yz} \\ \gamma_{xz} \\ \gamma_{xy} \end{bmatrix}, (1)$$

where $\sigma_{ij}$ denotes engineering stress, $\epsilon_{ij}$ denotes engineering strain, $\tau_{ij}$ denotes shear stress, $\gamma_{ij} = 2\,\epsilon_{ij}$ denotes shear strains, the subscripts $x$, $y$ and $z$ refer to the Cartesian axes and $G$, $\mu$, $\lambda$ and $\delta$ are four independent elastic constants. In previous studies [40], [60], we have demonstrated that $\delta$, which accounts for tissue tensile anisotropy, cannot be determined from guided wave propagation but that the in-plane Young's modulus can be approximated as $E_T = E \cong 3\mu$ assuming tensile isotropy ($\delta = 0$). Thus, among the four elastic constants, only $G$ and $\mu$, respectively the out-of-plane and in-plane shear moduli, are needed to predict corneal deformation under mechanical loading.

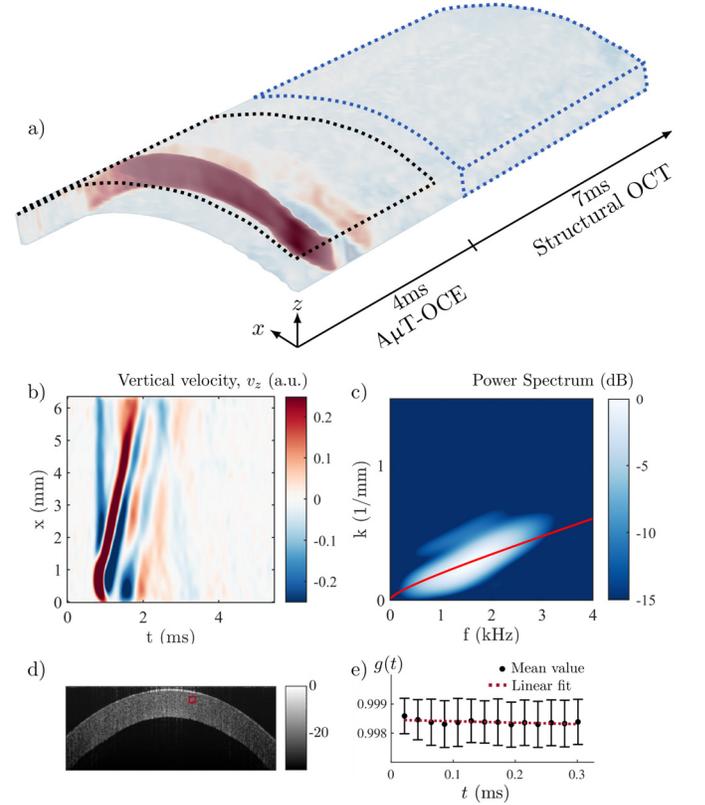

Figure 2. Stiffness reconstruction and depth penetration measurements using spectral domain time-resolved OCT. Pre-CXL measurements. a) 3D ($x, z$ and $t$), wavefield after AµT excitation. The top surface wavefield of the initial time sequence (black dotted region) is used for stiffness reconstruction and data at the end of the sequence are used for the phase decorrelation measurement. b) $x$-$t$ plot showing the top surface wavefield of the guided mode. c) $f$-$k$ spectrum obtained by 2D-FFT of the $x$-$t$-plot showing the dispersion signature of the first anti-symmetric mode $A_0$. The red curve indicates the best fit obtained with the NITI model. d) Structural OCT image obtained from averaging the last 7 ms of the raw OCT signal. e) Phase decorrelation function $g(\tau)$ at the location indicated by the red square on d).

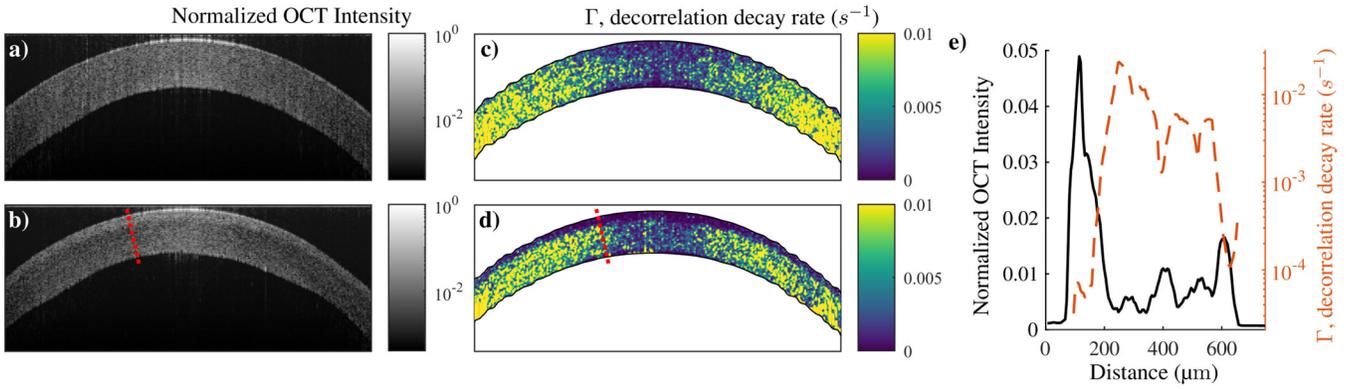

Figure 3. Short time decorrelation before and after CXL. Structural OCT images obtained from the last 7ms of the OCT scan for a) before CXL and b) after CXL. Maps of decorrelation coefficient $\Gamma$ for c) before and d) after CXL. e) Profile of OCT intensity and $\Gamma$ along the red dotted line shown in b) and d) for the CXL cornea.

The effects of CXL on the cornea depend on depth. Several recent studies showed that postoperative CXL corneas may experience non-uniform cross-linkage with depth. The transition between crosslinked (anterior) and non-crosslinked (posterior) parts tends to be sharp rather than smooth [61-63]. This effect is also observed in our experiments (see Fig. 3d, e). Thus, a two-layer model, although not exactly accurate, is considered a working model to quantify postoperative corneas. CXL was also shown to change collagen fiber diameter and interfibrillar spacing [61], but nothing suggests a modification of its macroscopic anisotropic organization.

Based on this observation, we developed a multi-layer model to predict wave propagation within CXL corneas (Supplementary Material 1) that accounts for any arbitrary number of layers, each with a stress-strain relationship given by Eq. (1) and linked by a perfect solid-solid boundary condition (continuity of normal components of stress and displacement across every interface). Accounting for the appropriate external boundary conditions (liquid below and air above the cornea) and the finite thickness of the medium, the dispersion relation for guided waves can be determined directly from stiffness moduli $G_n$ and $\mu_n$ and the thickness $h_n$ of each layer. A detailed description of the multi-layer model is provided in Supplementary Material 1. Although only 2 layers were considered in this study, the multi-layer model will allow further refinements if, for example, the transition between crosslinked and non-crosslinked areas needs a more accurate description.

Note that in an untreated cornea, only the first anti-symmetric mode, referred to as $A_0$, propagates in the range of frequencies that can be recorded in elastography (typically $< 5$ kHz). Because a CXL cornea is approximated as two horizontally assembled layers, each having a vertically aligned symmetry axis, this symmetry holds for the global material. Based on previous experiments on CXL corneas [38], [59], and symmetry conservation, the $A_0$-mode is expected to contain the primary energy of guided waves in CXL-treated tissue.

### D. Fitting pre- and post-CXL

The experimental $f$-$k$ spectrum (see Fig. 2c) was obtained by computing the 2D FFT of the $x$-$t$ plot. The shear moduli $G$ and $\mu$ in pre-CXL cornea were obtained from fitting the measured $f$-$k$ spectrum with the analytical dispersion relationship of the $A_0$ mode.

Prior to treatment, the cornea was assumed homogeneous, which in our model corresponds to a single layer bounded above by air and below by water. Using the model described above, the dispersion curve of a single layer material was used to determine the untreated properties of the cornea.

In CXL cornea, the thickness of both layers can be measured (see Section IIE) and the posterior layer can be assumed to still possess the original (*i.e.* untreated) elastic properties. Thus, the 2-layer model with known elastic moduli of the bottom (untreated) layer can be used to determine the stiffness of the top layer.

Since the $A_0$-mode carries depth averaged properties of the material, we also used a single-layer model to estimate the 'effective' stiffness of the treated cornea by fitting its $f$-$k$ spectrum with the analytical dispersion relationship of the $A_0$ mode computed for a single layer.

To ensure reliable fitting for all cases, we computed a goodness of fit metric $\Phi = \frac{\sum_f \chi_{fit}(f)}{\sum_f \chi_{max}(f)}$, where $\chi_{fit}(f)$ corresponds to the energy of the 2D spectrum covered by the best analytical solution (one or two layers) at a given frequency $f$ and $\chi_{max}(f)$ is the unconstrained maximal energy of the spectrum at frequency $f$. Based on recent results (see Supplemental Material in [38]), reliable fitting in human *ex vivo* corneas is associated with values of $\phi > 0.9$. An example of a 2D-spectrum and the fitted $A_0$ mode obtained with this procedure for the untreated case is shown in Fig. 2c. More details about the fitting procedure, including how to determine uncertainty intervals for $G$ and $\mu$ for both treated and untreated cases, are given in Supplementary Material 2.

### E. FEM simulations

We designed finite element (FEM) simulations in OnScale to determine the efficiency of our multi-layer NITI model for reconstruction of stiffness along depth. The geometry is shown in Fig. 4a. Corneal boundary conditions were replicated so that the material is bounded above by air and below by water. The speed of sound in all layers (material and water) was fixed to avoid reflection of compressional waves at air-tissue boundaries. It also improved the absorption of waves at the absorbing boundaries and, thus, avoided divergence of the simulations. We used transient excitation mimicking AμT experiments to generate broadband elastic waves within the material. More details about the simulations can be found in [42,50]. Based on phase-decorrelation measurements (see Section IIIF), we assumed that after CXL two layers with distinct thicknesses were formed within the cornea, the top layer being stiffer than the bottom one. The stiffness values assessed from experiments were also used in the simulations. We focused on the top surface signature of the simulated wave (see Fig. 4b) and its associated $f$-$k$ spectrum (see Fig. 4c) to study the efficiency of our fitting routine to quantify corneal elastic moduli and their variation with depth.

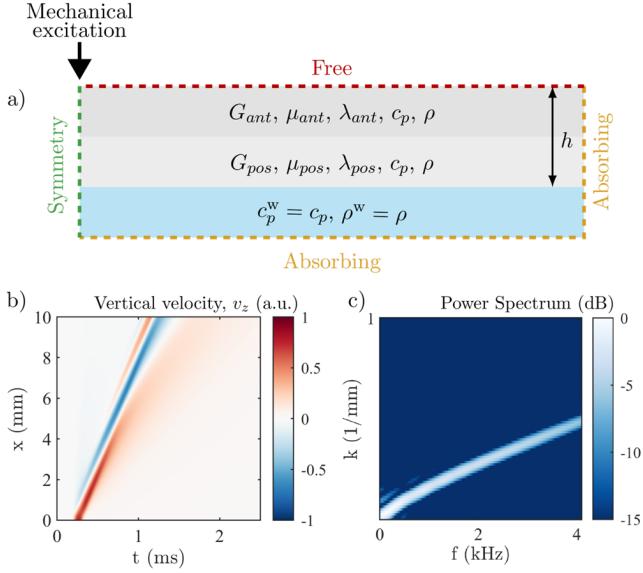

Figure 4. Finite element simulations to study the effects of a layered structure for a CXL cornea. a) Geometry of the two-layered material used in simulations, bounded above by air and below by water. b) Top surface spatio-temporal signature ($x$-$t$ plot) of the guided wave for the two-layer case with $G_{ant}$ = 296.8 kPa, $\mu_{ant}$ = 346 MPa, $G_{pos}$ = 59.5 kPa, $\mu_{pos}$ = 7.3 MPa, $h_{ant}$ = 150 μm and , $h_{pos}$ = 370 μm. d) 2D Fourier spectrum of the wave studied in b) showing the main propagating $A_0$-mode.

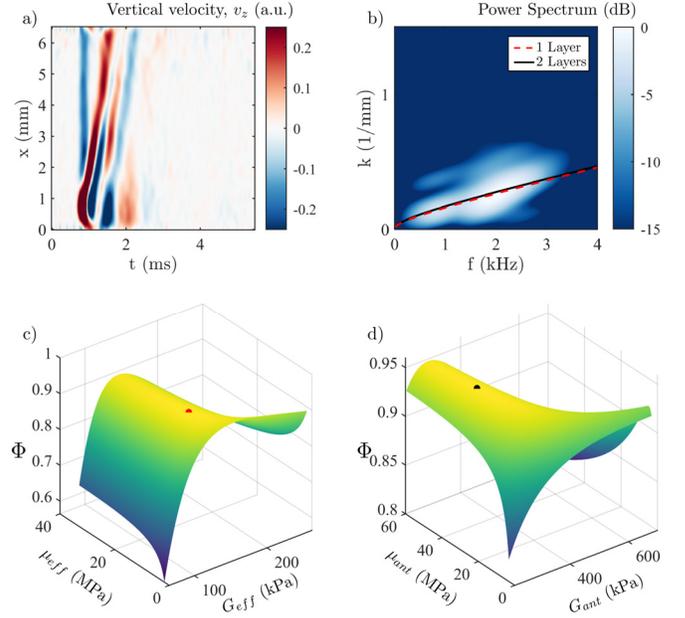

Figure 5. Fitting after crosslinking. a) Measured vertically polarized top-surface signature of the guided wave in the treated cornea. b) 2D-spectrum computed from a). c) Two-dimensional Goodness of Fit surface when the fit is performed with the single layer model (effective moduli fitting). d) Two-dimensional Goodness of Fit surface when the fit is performed with the two-layer model (top layer moduli fitting). For both c) and d), the optimum is indicated by a circular marker.

### F.  Phase Decorrelation OCT (PhD-OCT)

Blackburn *et al.* [64] have recently introduced a novel metric to track CXL penetration within the cornea using time-resolved OCT. It was shown that the phase decorrelation decay rate of the complex OCT signal is reduced in CXL areas and can be used to distinguish treated and untreated areas after the procedure.

In our study, the autocorrelation function of the signal $g(\tau)$ was computed over 15 consecutive samples at 46,500 Hz for six consecutive pixels within a given A-line:

$$g(\tau) = \left\langle \frac{\langle E(t)\, E^{*}(t+\tau)\rangle_{pixels}}{\sqrt{\langle E(t)\, E^{*}(t)\rangle_{pixels}} \times \sqrt{\langle E(t+\tau)\, E^{*}(t+\tau)\rangle_{pixels}}} \right\rangle, \quad (2)$$

which is expected to follow an exponential decay [67]:

$$g(\tau) = e^{-\Gamma \cdot \tau} \approx 1 - \Gamma \cdot \tau, \quad (3)$$

where $\Gamma$ is the decorrelation coefficient that is inversely proportional to the Brownian diffusion coefficient [67], meaning that the more coherent the material, the smaller the decorrelation coefficient. The procedure was performed starting at $n$, $n+1$, $n+2$, … A-lines, where $n$ is the first time-sample used for phase-decorrelation ($t(n) = 4$ ms). The decorrelation coefficient $\Gamma$ was then computed using the averaged $g(\tau)$ over the number of starting points by fitting with a first order polynomial (see Fig. 2e): $\langle g(\tau)\rangle = b - \Gamma \cdot \tau$, where $\langle\ \rangle$ denotes the average over the number of starting points. In crosslinked regions of the cornea (anterior), tissue stiffens, which implies that $\Gamma$ should be smaller than in the untreated region (posterior). For post-processing, we rejected all fits with b < 0.95, corresponding in general to peripheral regions where the signal to noise ratio (SNR) was too low.

**Table 1. Measured effective stiffnesses pre- and post-CXL.**

|  | Thickness, $h$ (μm) | Out-of-plane shear modulus, G (kPa) | In-plane shear modulus, $\mu$ (MPa) | Goodness of fit, $\phi$ |
|---|---|---|---|---|
| **Before CXL** | 575 | 59.5 ∓ (7, 12.8) | 7.6 ∓ (4.4,10) | 0.961 |
| **After CXL** | 520 | 127.5 ∓(14.3, 23.4) | 9.3 ∓(8, 17.2) | 0.953 |

## III. RESULTS

### A. Thickness of CXL layer

The spatial distribution of the OCT intensity signal (Figs. 3a, b, e) and phase decorrelation images (Figs. 3c, d, e) both show a clear layering effect in the treated cornea. The effect of CXL is not homogeneous across the cornea, with a more pronounced effect at the center (∼ 150 μm) than at the periphery (∼ 100μm). For the present case, we estimated that about 30% of the cornea was treated efficiently. As shown in Table 1, the treated cornea is thinner than that prior to CXL (its thickness reduced from 575 μm to 520 μm), as generally observed in the literature [68], [69].

### B. Stiffness of CXL-treated corneal layer

AμT-OCE scans, taking approximately 3 s to acquire and save data, were acquired before and after CXL. The signature of the vertical particle velocity (see Fig. 2ba) in the untreated cornea was used to compute the $f$-$k$ spectrum (see Fig. 2cb), which was fitted using the procedure detailed in Section IIC and Supplementary Material 2 assuming the cornea as a single homogeneous layer. The fitting routine was also detailed in our recent work [38], [59]. Results for the reconstructed in- ($\mu$) and out-of-plane ($G$) corneal shear moduli are shown in Table 1. It is worth mentioning that error bars are asymmetric, and the uncertainties in $\mu$ are always broader than those for $G$. More details about the procedure used to compute the uncertainty intervals are given in Supplementary Material 2 (see also [38], [59]).

To reconstruct the depth-dependent stiffness moduli of the cornea after it was treated with CXL, it is assumed that: i) the thickness of both layers can be measured using dynamic OCT from phase-decorrelation or intensity variation methods (see Fig. 3) which were estimated as about $h_{ant} = 150$ μm and $h_{pos} = 370$ μm with both methods; ii) the stiffness of the posterior layer is unchanged after CXL; iii) the effect of CXL is homogeneous in the anterior layer. Fixing known parameters (untreated cornea thicknesses and posterior stiffness moduli), stiffness moduli of the anterior cornea layer can be determined by fitting the wave dispersion curve in the $f$-$k$ domain with the 2-layer model (Fig. 5b, Supplementary Material 2). The goodness of fit surface plots are shown in Fig. 4c, d. We found an increase in both the stiffness moduli $G_{ant} = 296.8$ ∓ (26, 60) kPa and $\mu_{ant} = 34.6$ ∓ (16, 31) MPa compared to that for the posterior region $G_{post} = 59.5$ ∓ (7, 12.8) kPa and $\mu_{post} = 7.6$ ∓ (4.4, 10) MPa. The goodness of fit for the 2-layer model remained high, 0.95, which is still within the range of reliable fitting. These results are in good agreement with our recent study of corneas cross-lined throughout depth [38].

### C. Mixing rules of the mechanical moduli for layered materials

A theory for effective moduli of multi-layer materials was developed in the early 1970's for composite materials. It is now accepted in material science and broadly used in the development of composite structures [70]–[72]. For partially cross-linked cornea, it would be particularly interesting to have a rule to compute 'effective' moduli to estimate its ultimate stiffness, assess treatment efficiency, and predict corneal behavior and deformation post-operatively. The derivation of effective material moduli is based on 'mixing rules' of mechanical moduli across depth using the following assumptions: i) out-of-plane stresses and in-plane strains are uniform across thickness; ii) in-plane stresses and out-of-plane strains are averaged across thickness based on layer volume fractions. Note that the solution is valid only for very low frequencies. Sun *et al.* [71] have demonstrated that the effective low-frequency out-of-plane modulus $G_{eff}$ can be computed using the inverse rule of mixture for out-of-plane material constants of individual material layers.:

$$G_{eff} = \left(\sum_n \frac{h_n/h}{G_n}\right)^{-1}, \qquad (4)$$

where $h$ is the total material thickness, $h_n$ is the thickness of the $n^{th}$ layer and $G_n$ is the out-of-plane modulus of the $n^{th}$ layer. On the other hand, the effective low-frequency in-plane modulus $\mu_{eff}$ can be obtained from the rule of 'mixtures':

$$\mu_{eff} = \sum_n \mu_n \cdot \frac{h_n}{h}, \qquad (5)$$

where $\mu_n$ is the in-plane modulus of the $n^{th}$ layer. Based on the mixture rules described above, the effective low-frequency moduli of a partially treated cornea can be computed, and in our case are equal to $G_{eff} = 77.3$ kPa and $\mu_{eff} = 15.4$ MPa.

### D. Effective 'guided wave' corneal moduli

As noted above, elastic moduli determined for the anterior (CXL-treated) corneal layer can be used to compute effective corneal mechanical moduli, with $\mu_{eff}$ calculated using a simple rule of mixture, whereas $G_{eff}$ required the inverse rule of mixture.

It was interesting for us to check if this model could also describe guided wave behavior in the partially crosslinked cornea when considered as an effective homogeneous material. Using our analytical model, we studied the accuracy of this averaging method to predict effective 'guided wave' moduli, *i.e.* a pair of $\mu_{guided}$ and $G_{guided}$ that would fit the analytical $A_0$-mode dispersion curve of the N-layer case as if the $A_0$-mode dispersion curve is obtained for a single layer with averaged properties computed from Eqs. (4) and (5). The results are presented in Fig. 6 for a material with a total thickness $h = 500$ μm, made of two layers with $G_{ant} = 300$ kPa and $\mu_{ant} = 30$ MPa, and $G_{pos} = 60$ kPa and $\mu_{pos} = 5$ MPa. We studied different repartitions of the layers with either 30% (Fig. 6a), 50% (Fig. 6b) or 70% (Fig. 6c) of the total thickness for the anterior layer respectively. The $A_0$-mode computed with the effective mechanical moduli using the mixture rules described above does not match the exact analytical solution computed using the individual mechanical moduli of the layers, *i.e.* using the 2-layered model directly. The difference is especially pronounced in the high-frequency range. Thus, the mixture rules describing effective mechanical moduli in the cornea (or any other layered medium) cannot describe guided wave behavior in the effective medium. This finding is very important and shows that the model of an effective medium cannot be used to describe guided wave behavior. Therefore, reconstruction of effective mechanical moduli from OCE measurements in the cornea should be performed with care.

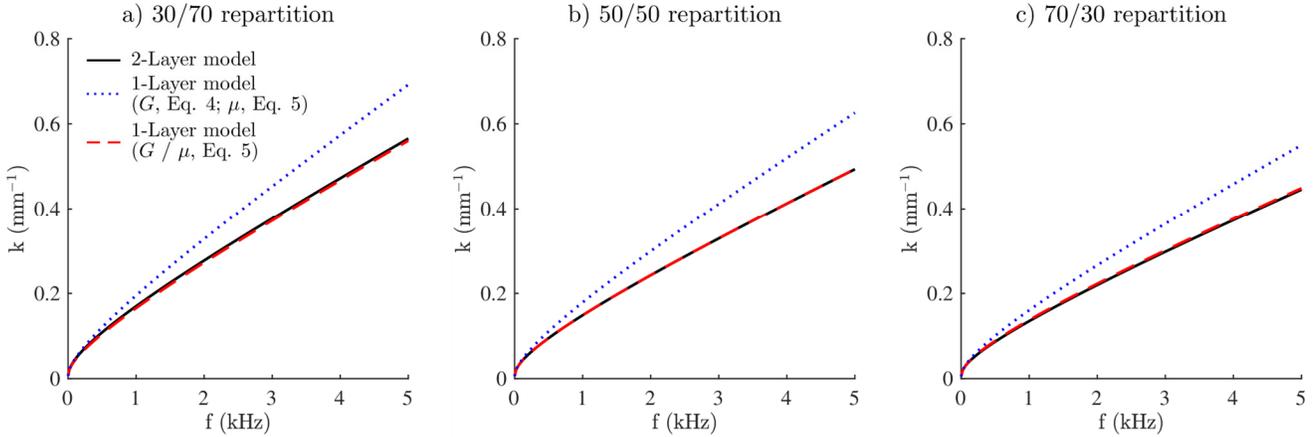

Figure 6. Study of effective 'guided wave' shear moduli and 'effective engineering' shear moduli for the $A_0$-mode dispersion curve. Three analytical solutions are plotted, the exact one obtained from the N-Layer model (continuous line); the effective one associated with the mixing rules introduced in equations (4) and (5) (dotted line); an additional one where both effective moduli are computed from an arithmetic weighted average (dashed line). The three cases (a), (b) and (c) are obtained for two layers, where the moduli of the top one are $G_{ant} = 300$ kPa and $\mu_{ant} = 30$ MPa and for the bottom layer are $G_{pos} = 60$ kPa and $\mu_{pos} = 5$ MPa. The total thickness is $h = 500$ µm, but the distribution of the layers varies so that: a) the top layer takes 30% of the total thickness and the bottom one 70%; b) both top and bottom layers take 50% of the total thickness; c) the top layer takes 70% of the total thickness and the bottom one 30%.

The second question is which mixture model would best describe guided wave behavior in a partially CXL treated cornea. This is an open and non-trivial question that is outside the scope of this paper. However, as shown in Fig. 6, the use of a simple direct mixture rule for both in- and -out-of-plane moduli (Eq. (5)) produces a dispersion curve closely matching the 2-layer model. In particular, equal thicknesses layers (50/50 split ratio, Fig. 6b) show a near-perfect match. The larger the difference between the layer thicknesses, the larger the difference between the solutions, although the difference remains small. We simulated many different cases, varying thicknesses, moduli, and the number of layers, and can conclude that the mixture rule described by Eq. (5) for both moduli gives a dispersion solution very close to the exact solution. We can treat this observation as an experimental fact but, unfortunately, we do not have its rigorous analytical confirmation.

Based on the observation above, we also reconstructed elastic moduli in the anterior (CXL treated) layer of the cornea using Eq. (5) as a mixture rule for both in- and out-of-plane moduli and obtained $G_{guided} = 127.9$ kPa and $\mu_{guided} = 15.4$ MPa, which is in overall good agreement with the directly (using 2-

layer model) measured post-CXL effective moduli $G_{eff} = 127.5 \mp (14.3, 23.4)$ kPa and $\mu_{eff} = 9.3 \mp (8, 17.2)$ MPa. Again, we do not have a mathematically rigorous solution for the 'effective' guided wave behavior but can confirm that both theoretical simulations and experimental results suggest a simple mixture rule for both mechanical moduli to describe the effective guidance of mechanical waves in multi-layered nearly-incompressible media in general, and in the cornea in particular.

## IV. DISCUSSION AND CONCLUSIONS

In this study we combined structural OCT with dynamic AµT-OCE to assess the penetration depth of CXL treatment in the cornea. Analyzing the brightness of structural OCT images and the rate of image decorrelation between consecutive B-scans, we can conclude that there is a sharp transition between CXL and untreated cornea layers. This finding allowed us to measure the thickness of the CXL treated layer and suggested a model of a two-layer medium to reconstruct both in- and out-of-plane elastic moduli in this layer.

Using AµT-OCE, we tracked guided wave propagation in a cornea before and after CXL. Using an appropriate NITI model (in each layer), we quantified the localized stiffening of cornea with CXL. Because the $A_0$-mode occupies the whole thickness, it carries averaged information about material stiffness (*i.e.,* averaged over its two parts). We proposed two ways to determine the effective moduli in the treated cornea. In the first case, the 2D-spectrum in the post-CXL cornea was fitted with the two-layer model with known thicknesses of each layer and known moduli of the untreated part (obtained from OCE measurements in the untreated cornea). The effective mechanical moduli of the entire cornea can be then calculated using Eqs. (4) and (5) respectively for out-of- and in-plane moduli. The second way assumed the effective guided wave behavior in the cornea, where the 2D-spectrum post-CXL is fitted with a single layer model.

We found that the mixture rule for layered materials in continuous mechanics (see Eqs. (4) and (5)) works differently for in- and out-of-plane moduli. From these expressions the effective moduli of partially CXL-treated cornea can be computed, which may be important in assessing CXL surgery and predicting its outcome. However, the same mixture rules cannot be applied to quantify effective guided wave behavior in the layered medium. Both numerical simulations and *ex vivo* experiments in the human cornea sample suggest a simple mixture rule (Eq. (5)) for both corneal moduli, but we do not have a rigorous analytical confirmation of this observation.

One of the limitations of the study is that it assumes the post-CXL cornea consists of two homogeneous layers. This model was found reasonable in several previous studies. In fact, penetration of the riboflavin/dextran solution into the cornea and the unequal UVA irradiation with depth might suggest a more gradual transition of stiffness between layers. A multi-layer model following the approach described here can be used to compute the $A_0$-mode dispersion curve for an arbitrary number of layers, their stiffnesses and thicknesses and, thus, quantify CXL treatment outcomes with better precision.

Recent results suggest that reverberant OCE can reconstruct the stiffness depth-dependent variation [29]. It would be interesting to compare our method with reverberant OCE, which will be the object for our future studies. Note, however, that reverberant OCE is not currently feasible *in vivo* because it uses contact vibrators. This is why guided wave-based OCE is still the only method capable of *in vivo* non-contact measurements of corneal anisotropic elasticity, ultimately to be spatially resolved.

Finally, we have shown that phase-sensitive OCT combined with AμT wave excitation can accurately measure both the structure of human corneas and the depth-dependence of moduli due to crosslinking. These findings are essential for building personalized models of corneal deformation following CXL, and thus better adapt crosslinking therapy for clinical use and predict its outcomes. However, further experiments on a larger group of samples are required to generalize the present results.

# Supplementary Material

---

# Possible depth-resolved reconstruction of shear moduli in the cornea following collagen crosslinking (CXL) with optical coherence tomography and elastography.

Gabriel Regnault, Mitchell A. Kirby, Ruikang K. Wang, Tueng T. Shen, Matthew O'Donnell, and Ivan Pelivanov

**Supplementary Material 1**: **Analytical solution for a multi-layer nearly incompressible transverse isotropic (NITI) medium.**

Assume that CXL corneas can be modelled as a laminate. Each layer has finite thickness and defined in- and out-of-plane elastic moduli. The first layer is bounded on the top by air and the last layer is bounded on the bottom by a liquid to mimic corneal in vivo conditions. Because collagen fibers are oriented randomly in the equatorial plane, it possesses a symmetry across fibers, *i.e.* in the direction normal to the corneal surface, which is mathematically described as transverse isotropy. In addition, like most biological tissues, the cornea is nearly incompressible. In our recent study [1], we presented a model of guided wave propagation in a NITI medium and showed that it can be used to determine both in- and out-of-plane shear moduli on both *ex vivo* and *in vivo* rabbit, porcine and human cornea models [1–4]. In this section, we present an extension of the NITI model accounting for multiple cornea layers.

## a) Dimensionless equations of motion

Because the macroscopic symmetry of the cornea is not changed following crosslinking, we assume that it consists of at least two layers of finite thickness with distinct in- and out-of-plane shear moduli $\mu$ and $G$ respectively. The density of every layer, $\rho$, is assumed to be identical for all layers and equal to that of the liquid bounding the lower layer, $\rho_l = \rho = 1000 \ \text{kg·m}^{-3}$. Using Voigt's notation, the stress-strain relationship in each layer can be written as:

$$\begin{bmatrix} \sigma_{xx} \\ \sigma_{yy} \\ \sigma_{zz} \\ \tau_{yz} \\ \tau_{xz} \\ \tau_{xy} \end{bmatrix} = \begin{bmatrix} \lambda + 2\mu & \lambda & \lambda & & & \\ \lambda & \lambda + 2\mu & \lambda & & & \\ \lambda & \lambda & \lambda + \delta & & & \\ & & & G & & \\ & & & & G & \\ & & & & & \mu \end{bmatrix} \begin{bmatrix} \epsilon_{xx} \\ \epsilon_{yy} \\ \epsilon_{zz} \\ \gamma_{yz} \\ \gamma_{xz} \\ \gamma_{xy} \end{bmatrix}, \qquad \text{(S1)}$$

where $\lambda = \rho c_p^2 - 2\mu$, and $c_p$ is the speed of the longitudinal wave that ensures incompressibility of the material (the Poisson's ratio of each layer taken individually is $\nu \sim 0.5$).

Newton's second law yields the wave equation of motion in terms of the displacement vector $\vec{u} = (u, v, w)$,

$$\frac{\partial \sigma_{ij}}{\partial x_j} = \rho \frac{\partial u_i}{\partial t^2}, \qquad \text{(S2)}$$

Because in our experiments an AµT source launches a pseudo-line source, we can assume a plane-strain state (no displacement polarized along the $y$-axis: $v = 0$, and no propagation along the $y$-axis: $u = u(x, z), w = w(x, z)$). As such, in the NITI model, the equations of motion can be expressed:

$$\rho . u_{tt} = (\lambda + 2\mu)u_{xx} + G u_{zz} + (\lambda + G)w_{xz} ,\qquad \text{(S3)}$$
$$\rho . w_{tt} = G w_{xx} + (\lambda + 2\mu)w_{zz} + (\lambda + G)u_{xz} ,\qquad \text{(S4)}$$

where the lower indexes indicate derivatives with time ($t$) or spatial coordinates (x, z).

By introducing the scales:

Position: $x \sim H = \sum_i h_i$ ,
Displacement: $u \sim H$ ,

Time: $t \sim H.\sqrt{\frac{\mu_M}{\rho}}$ , $\mu_M = \max(\mu_i)$ ,

Frequency: $f \sim \sqrt{\frac{\mu_M}{\rho}}.\frac{1}{H}$ ,

Wavenumber: $k \sim \frac{1}{H}$ ,

we can define the dimensionless parameters $t^* = \frac{t}{H}\sqrt{\frac{\mu_M}{\rho}}$, $u^* = \frac{u}{H}$, $x^* = \frac{x}{H}$, $f^* = fH\sqrt{\frac{\rho}{\mu_M}}$, $k^* = kH$.

By applying this change of variables, we have:

$$\frac{\partial u}{\partial x} = \frac{\partial H \times u^*}{\partial x^*} \times \frac{\partial x^*}{\partial x} = H \times \frac{\partial u^*}{\partial x^*} \times \frac{1}{H} = \frac{\partial u^*}{\partial x^*} ,\qquad \text{(S5)}$$

$$\frac{\partial^2 u}{\partial x^2} = \frac{\partial}{\partial x}\left(\frac{\partial u^*}{\partial x^*}\right) = \frac{1}{H}\frac{\partial^2 u^*}{\partial x^{*2}} ,\qquad \text{(S6)}$$

$$\frac{\partial^2 u}{\partial t^2} = \frac{\mu_M}{\rho H}\frac{\partial^2 u^*}{\partial t^{*2}} ,\qquad \text{(S7)}$$

which eventually leads to the dimensionless equations of motion:

$$\rho . u_{tt}^* . \frac{\mu_M}{\rho H} = (\lambda + 2\mu)u_{xx}^*.\frac{1}{H} + G u_{zz}^*.\frac{1}{H} + (\lambda + G)w_{xz}^*.\frac{1}{H} ,\qquad \text{(S8)}$$

$$\rho \mu_M . w_{tt}^* \frac{\mu_M}{\rho H} = G w_{xx}^*.\frac{1}{H} + (\lambda + 2\mu)w_{zz}^*.\frac{1}{H} + (\lambda + G)u_{xz}^*.\frac{1}{H} ,\qquad \text{(S9)}$$

and after rearranging:

$$u_{tt}^* = \beta^2 u_{xx}^* + \alpha^2 u_{zz}^* + \gamma^2 w_{xz}^* ,\qquad \text{(S10)}$$
$$w_{tt}^* = \alpha^2 w_{xx}^* + \beta^2 w_{zz}^* + \gamma^2 u_{xz}^* ,\qquad \text{(S11)}$$

with

$$\alpha^2 = \frac{G}{\mu_M} ,\qquad \text{(S12)}$$

$$\beta^2 = \frac{(\lambda + 2\mu)}{\mu_M} ,\qquad \text{(S13)}$$

$$\gamma^2 = \frac{(\lambda + G)}{\mu_M} .\qquad \text{(S14)}$$

For the sake of simplicity, we will later omit the $*$ symbol to refer to dimensionless variables.

### b) Dispersion relationship of guided waves in a multilayered NITI material

We consider here a material made of multiple layers with identical density, each of finite thickness $h_i$. As such, we have a system of $[2 \times N]$ equations for the $N$ layers:

$$u_{i,tt} = \beta_i^2 u_{xx} + \alpha_i^2 u_{zz} + \gamma_i^2 w_{xz} \tag{S15}$$

$$w_{i,tt} = \alpha_i^2 w_{xx} + \beta_i^2 w_{zz} + \gamma_i^2 u_{xz} \tag{S16}$$

$$\alpha_i^2 = \frac{G_i}{\mu_N} \tag{S17}$$

$$\beta_i^2 = \frac{(\lambda_i + 2\mu_i)}{\mu_N} \tag{S18}$$

$$\gamma_i^2 = \frac{(\lambda_i + G_i)}{\mu_N}. \tag{S19}$$

We assume harmonic solutions of the form:

$$u_i(x,z,t) = A_i e^{i(kx + l_i z - \omega t)}, \tag{S20}$$

$$w_i(x,z,t) = B_i e^{i(kx + l_i z - \omega t)}. \tag{S21}$$

Without loss of generality, we assume $B_i = 1$. By substituting equations (S20) and (S21) into the equations of motion (S15) and (S16), we can determine for each frequency and wavenumber the constants $l_i$ and $A_i$ so that:

$$l_i = \pm \sqrt{\frac{1}{2}\left[\phi_i \pm \sqrt{\phi_i - 4q_{\alpha_i}^2 q_{\beta_i}^2}\right]}, \tag{S22}$$

$$A_i = \pm\left[-\frac{\frac{\sqrt{2}\gamma_i^2 k}{\alpha_i^2}\sqrt{\phi_i \pm \sqrt{\phi_i - 4q_{\alpha_i}^2 q_{\beta_i}^2}}}{\phi_i + \frac{2\beta_i^2}{\alpha_i^2}q_{\beta_i}^2 \pm \sqrt{\phi_i - 4q_{\alpha_i}^2 q_{\beta_i}^2}}\right], \tag{S23}$$

where

$$\phi_i = \frac{\gamma_i^4 k^2}{\alpha_i^2 \beta_i^2} - \frac{\alpha_i^2}{\beta_i^2}q_{\alpha_i}^2 - \frac{\beta_i^2}{\alpha_i^2}q_{\beta_i}^2, \tag{S24}$$

$$q_{\alpha_i}^2 = k^2 - \frac{\omega^2}{\alpha_i^2}, \tag{S25}$$

$$q_{\beta_i}^2 = k^2 - \frac{\omega^2}{\beta_i^2}. \tag{S26}$$

The full solutions for a given layer can be expressed as the combination of 4 partial waves:

$$u_i(x,z,t) = \sum_{j=1}^{4} C_{i,j} A_{i,j} e^{i l_{i,j} z} e^{i(kx - \omega t)}, \tag{S27}$$

$$w_i(x,z,t) = \sum_{j=1}^{4} C_{i,j} e^{i l_{i,j} z} e^{i(kx - \omega t)}. \tag{S28}$$

In the fluid, the dimensionless velocity potential is:

$$\Phi = C_{N,5} e^{\epsilon z} e^{i(kx - \omega t)}, \tag{S29}$$

where $\epsilon = \sqrt{k^2 - \frac{\omega^2}{\delta^2}}$ and $\delta^2 = \frac{\rho}{\mu_N}c_p^2$.

The constants $C_{i,j}$ are chosen so that the solutions satisfy the boundary conditions:

I.    Traction free air-solid interface

$$\sigma_{1,xz} = 0 \qquad \text{at} \qquad z_1 = 1 \,, \qquad\qquad (S30)$$

$$\sigma_{1,zz} = 0 \qquad \text{at} \qquad z_1 = 1 \,, \qquad\qquad (S31)$$

II.   Continuity of normal components of stress and displacement between each layer

$$\sigma_{i,xz} = \sigma_{i+1,xz} \quad \text{at} \qquad z_i = 1 - \frac{\sum_{k=1}^{i} h_k}{H} \,, \qquad\qquad (S32)$$

$$\sigma_{i,zz} = \sigma_{i+1,zz} \quad \text{at} \qquad z_i = 1 - \frac{\sum_{k=1}^{i} h_k}{H} \,, \qquad\qquad (S33)$$

$$u_i = u_{i+1} \qquad \text{at} \qquad z_i = 1 - \frac{\sum_{k=1}^{i} h_k}{H} \,, \qquad\qquad (S34)$$

$$w_i = v_{i+1} \qquad \text{at} \qquad z_i = 1 - \frac{\sum_{k=1}^{i} h_k}{H} \,. \qquad\qquad (S35)$$

III.  Medium-fluid boundary conditions (zero tangential stress, continuity of normal stress components and speed)

$$\sigma_{N,xz} = 0 \qquad \text{at} \qquad z_N = 0 \,, \qquad\qquad (S36)$$

$$\sigma_{N,zz} = \sigma_{zz}^{f} \qquad \text{at} \qquad z_N = 0 \,, \qquad\qquad (S37)$$

$$\dot{w}_N = \dot{w}^{f} \qquad \text{at} \qquad z_N = 0 \,. \qquad\qquad (S38)$$

Substituting the general solution into the boundary conditions yields a [4N+1 × 4N+1] homogeneous system for the coefficient: $\mathbf{Mc = 0}$. This system has a nontrivial solution if and only if the determinant of $\mathbf{M}$ is zero. For a given angular frequency $\omega$, the wavenumber $k$ associated with different wave types (pure shear or guided modes) can be found by minimizing the absolute value of the determinant.

### c) $A_0$-mode tracing for N-layers

Because each layer in the laminate has a vertical symmetry axis, this symmetry holds for the global material. Based on previous AµT-OCE experiments in CXL corneas [3,4], and accounting for symmetry conservation, it follows that the first anti-symmetric mode, $A_0$, carries moat of the wave energy. Using the analytical model detailed above, we can determine the $A_0$-mode dispersion curve for a material made of N layers with an arbitrary distribution of NITI elastic moduli and thickness.

To compute the $A_0$-mode of the global material, we compute the determinant of the matrix M for each frequency, $f$, and each interrogated wavenumber, $k$. One of the minima corresponds to the $A_0$-mode speed such that $c_{A_0}(f) = f / \min(k)$. Such a map is shown in figure S1 below. On top of the $A_0$ mode dispersion curve, we can see constant phase velocity curves, which corresponds to pure shear wave speeds $\left( C_{s,i} = \sqrt{\frac{G}{\rho}} \right)$ in every layer. Because the $A_0$-mode and the pure shear wave coexist in the numerical solution, the $A_0$-mode can be inaccurate in some situations, and their separation should be done with care.

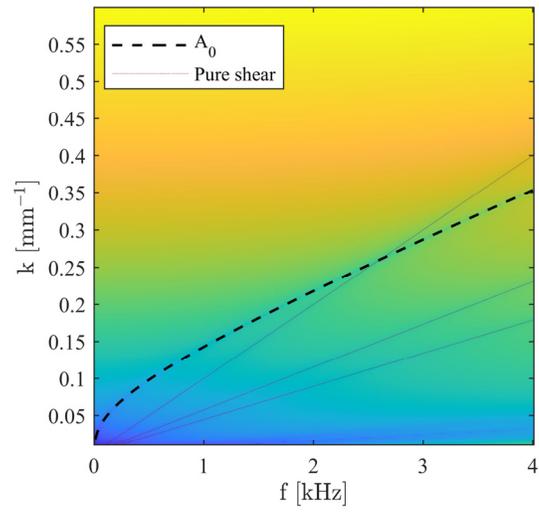

Figure S1. Determinant Matrix map for a given range of $f$ and $k$. The color-bar indicates the value of the determinant. This case corresponds to a total thickness of 0.5 mm, with uniform distribution of thickness among layers and stiffness moduli $G$ = [500 kPa, 300 kPa, 100 kPa] and $\mu$ = [10M Pa, 6 MPa, 2 MPa], respectively from top to bottom. The straight lines within the map correspond to the pure shear wave solution of the three individual layers.

<u>**Supplementary Material 2**</u>: **Fitting the measured $A_0$-mode dispersion curve.**

   *a) Rough estimation of the elastic moduli*

An iterative routine to estimate in-plane tensile and out-of-plane shear moduli ($\mu$ and $G$, respectively) was previously presented. The method finds the theoretical $A_0$ dispersion mode of a guided elastic wave in a nearly-incompressible transverse isotropic (NITI) material that most closely matches experimentally measured elastic wave propagation [1,2]. Iteration was performed in the frequency-wavenumber ($f$-$k$) domain using simplex optimization (*fminsearch*, MATLAB, MathWorks, Natick, MA), where the best-fit theoretical dispersion relation was performed by maximizing the following objective function:

$$\Phi(\mu, G) = \frac{1}{N_f} \sum_f \sum_k w(f, k; \mu, G) |\hat{v}(f, k)|^2 - \beta \left| \frac{\mu}{\lambda} \right|, \qquad (S39)$$

where $\hat{v}$ is the normalized 2D Fourier spectrum of the measured time-space distribution of vertical displacements induced by the propagating guided wave. The function $w(f, k; \mu, G)$ is related to the $A_0$-mode solution for a NITI ($G < \mu$) or isotropic ($G = \mu$) material. The regularization term, $\beta$, satisfying the nearly-incompressible assumption, was set to 1 based on an L-curve analysis.[2] The value of $\lambda$ was updated at each iteration according to $\lambda = \rho c_L^2 - 2\mu$. Note that prior to computing the 2D Fourier spectrum, a temporal super gaussian filter ($SG$) that followed the maximum vibration velocity of the wave-field $t_m^{wf}(x)$ at each discrete position $x$, was applied:

$$SG(t) = \exp\left[ -\left( \frac{1}{2} \left( \frac{t - t_m^{wf}(x)}{\sigma_t} \right)^2 \right)^2 \right], \qquad (S40)$$

where $\sigma_t = 0.5$ ms. This step was included to prevent error in the fitting routine due to low frequency noise induced by the experimental setup.

The optimization function was not posed as a least-squares error problem. Rather, the procedure sought to maximize the spectral energy contained in a small, weighted window around the $A_0$ mode. At a given frequency $f$ and current parameter set $(\mu, G)$, the dispersion relation solver returned the wavenumber associated with the $A_0$ mode, $k_0$. The weighted window function at each frequency centered at the corresponding $k_0$ was defined by $w(f, k; \mu, G)$, where $\Delta k$ was the discrete wavenumber sampling interval:

$$w(f, k; \mu, G) = \begin{cases} e^{-\frac{1}{2}\left( \frac{k - k_0(f, \mu, G)}{1.2\Delta k} \right)^2}, & |k - k_0(f, \mu, G)| \leq 3\Delta k \\ 0, & \text{otherwise} \end{cases}, \qquad (S41)$$

and the weights were assigned at each frequency with a peak value at the wavenumber of the theoretical dispersion curve with these weights decaying (with a Gaussian distribution) with distance from the theoretical curve.

The dispersion relation estimated the location of the $A_0$-mode (e.g. the wavenumber of the mode for a given frequency) but not the spectral energy in a frequency-wavenumber bin (which depends on the method used to excite guided waves). Essentially, this approach found the dispersion curve that overlapped with the maximum spectral power. Multiple physical parameters were fixed, including the corneal density ($\rho$ = 1000 kg/m$^3$), corneal longitudinal wave speed ($c_L$ = 1540 m/s), and mean corneal thickness (measured from B-mode OCT images). The cornea was assumed bounded from below by water with a density of 1000 kg/m$^3$ and longitudinal wave speed of 1540 m/s. To avoid convergence to a local (as opposed to global) maximum in equation S.1, five independent fits were performed, with quasi-random initial values of $G_0$ and $\mu_0$. The final output of $\mu$ and $G$ were set to those corresponding to the highest value in equation S39.

Note:

The same procedure was applied for the single layer and two-layer cases. For the latter, only the top layer moduli were fitted. To allow a better comparison, the regularization term for the two-layer case was computed using the weighted arithmetic average value of $\mu = \frac{h_{top}\mu_{top} + h_{bot}\mu_{bot}}{h_{top} + h_{bot}}$.

### b) *Fine search and goodness of fit*

To quantify how well the NITI solution fits experimental data, a goodness-of-fit (GOF) metric was introduced based on an "unconstrained, global optimum" for the objective function, $\Phi_{\max}$. At each frequency $f$, the following optimization problem was solved:

$$k_{\max} = \underset{\tilde{k}}{\mathrm{argmax}} \sum_k \tilde{w}(f, k; \tilde{k}, \mu, G) |\hat{v}(f, k)|^2 \, , \qquad (S42)$$

$$\tilde{w}(f, k; \tilde{k}, \mu, G) = \begin{cases} e^{-\frac{1}{2}\left(\frac{k - \tilde{k}}{1.2\Delta k}\right)^2}, & |k - \tilde{k}| \leq 3\Delta k \\ 0, & \text{otherwise} \end{cases} \, , \qquad (S43)$$

$$\Phi_{\max}(\mu, G) = \frac{1}{N_f} \sum_f \sum_k \tilde{w}(f, k; k_{\max}, \mu, G) |\hat{v}(f, k)|^2 \, . \qquad (S44)$$

This was equivalent to finding the Gaussian-weighted window containing the most spectral energy at each frequency independently and summing those contributions to the total mode energy. Since $k_{\max}$ may vary at each frequency independent of the dispersion relation, $k_{\max}$ may not follow the mode shape exactly and may not even be smooth. However, this means that $\Phi_{\max}$ represents an upper bound on the value of $\Phi$ calculated during fitting. We therefore produced the following GOF metric:

$$g_{\mathrm{NITI}} = \frac{\Phi_{\mathrm{NITI}}}{\Phi_{\max}} \, . \qquad (S45)$$

$\Phi_{\mathrm{NITI}}$ covered the maximum energy with unique combinations of $\mu$ and $G$, constrained by the shape of the $A_0$ dispersion curve in the NITI model, while $\Phi_{\max}$ captured the unconstrained energy. Thus, $g_{\mathrm{NITI}}$ indicated the portion of the maximum possible mode energy captured by a given $A_0$ dispersion curve. Values near $g_{\mathrm{NITI}} <\approx 1$ suggest that the theoretical dispersion curve is well described by the NITI model, where $\Phi_{\mathrm{NITI}}$ captures nearly all the measured mode's energy.

Based on this GOF metric, we proposed to refine the fitting by building lookup tables around the rough estimates of $G$ and $\mu$. We build a 2D-surface plot of the GOF metric $\Phi$, whose maximum indicates the pair of moduli that best fit the 2D-spectrum.

### c) *Quantifying uncertainty intervals*

While $g_{\mathrm{NITI}}$ provides an estimate for how well the theoretical $A_0$-mode matches experimental data, it alone does not provide confidence intervals on the output moduli $G$ and $\mu$. Due to the nature of maximization in the weighted fitting, residual errors are not generated, making traditional confidence interval methods difficult to apply. For example, a low value of $g_{\mathrm{NITI}}$ would suggest that the $A_0$ dispersion curve calculated from the NITI model poorly described the actual dispersion measured within the sample. In such a case, moduli estimates should have increased uncertainty.

As discussed in one of our previous studies [3], the variation of the GOF for an given sample at a fixed IOP is about 1%. We used this to build error bars for the present study. This is illustrated below in Fig. S2 with the dataset corresponding to the measurement before CXL. In this case, we measured $G = 59.5$ kPa, $\mu = 7.6$ MPa and a GOF of 0.961. First, we fitted the projection of the surface plot for a value of $g_{\mathrm{NITI}} = (1 - 0.01) \times \max(g_{\mathrm{NITI}})$ (*i.e.* 1% below the optimal GOF), with an ellipsoidal function that best describes the shape of the iso-goodness levels. Then, we computed the uncertainty intervals as the intersection of the horizontal and vertical direction with the fitted ellipse. This results in asymmetric uncertainty intervals, reflecting the asymmetric variation of the 2D GOF function.

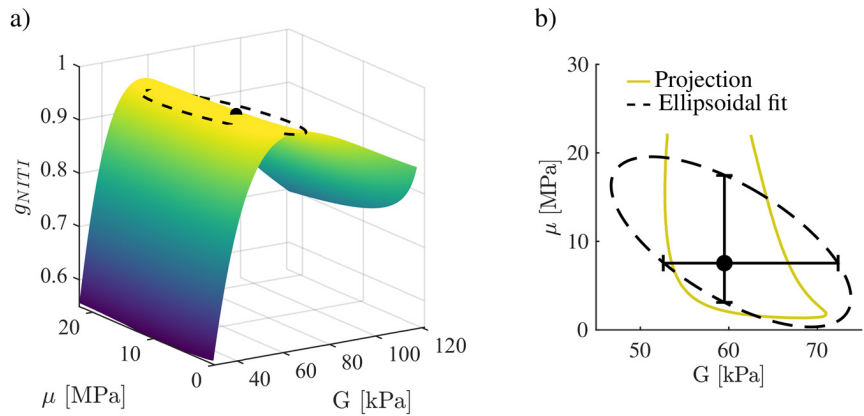

Figure S2. Estimation of uncertainty intervals. a) 2D surface plot of $g_{NITI}$. b) Projection of the surface plot for $g_{NITI} = (1 - 0.01) \times \max(g_{NITI})$. The error bars are computed as the intersection of the vertical and horizontal direction with the fitted ellipse.

<u>**Supplementary Material 3**</u>: Mixing rules for shear moduli in a layered material

Several groups have studied the effective mechanical properties of layered structures [5–7]. In this section, we reproduce the general steps to compute effective material properties.

Consider a material made of consecutive laminas, whose stiffness and principal orientations are known (see Figure S3). Let's further assume that all layers have the same principal orientation. Note that this last assumption is not necessary but simplifies the calculations and corresponds to our situation.

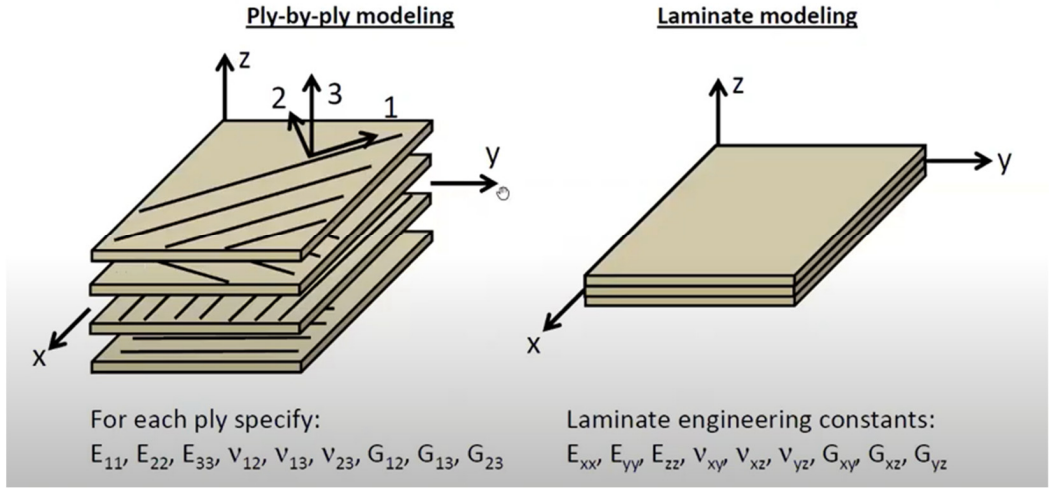

Figure S3. Schematic of laminas and laminated system under study. Reproduced with permission from [8].

By definition, we know for each lamina the constitutive equation that relates stress and strain:

$$\boldsymbol{\sigma}^k = \boldsymbol{C}^k \boldsymbol{\epsilon}^k, \tag{S46}$$

where superscript "k" indicates a given laminate, or in Voigt's notation:

$$
\begin{bmatrix}
\sigma_x^k \\
\sigma_y^k \\
\sigma_z^k \\
\sigma_{xz}^k \\
\sigma_{yz}^k \\
\sigma_{xy}^k
\end{bmatrix}
=
\begin{bmatrix}
\lambda^k + 2\mu^k & \lambda^k & \lambda^k & 0 & 0 & 0 \\
\lambda^k & \lambda^k + 2\mu^k & \lambda^k & 0 & 0 & 0 \\
\lambda^k & \lambda^k & \lambda^k + 2\mu^k & 0 & 0 & 0 \\
0 & 0 & 0 & 2G^k & 0 & 0 \\
0 & 0 & 0 & 0 & 2G^k & 0 \\
0 & 0 & 0 & 0 & 0 & 2\mu^k
\end{bmatrix}
\begin{bmatrix}
\epsilon_x^k \\
\epsilon_y^k \\
\epsilon_z^k \\
\epsilon_{xz}^k \\
\epsilon_{yz}^k \\
\epsilon_{xy}^k
\end{bmatrix}. \tag{S47}
$$

The idea is to determine an effective stiffness matrix $\overline{\boldsymbol{C}}$ associated with the global material. To do so, five main assumptions are made:

i) Following the assumption of continuity of the z-components of stress at the layer's interfaces, the out-of-plane stresses are assumed to be uniform along the thickness, and thus equal the to the out-of-plane stress within the overall material (subscript $\overline{\phantom{-}}$:

$$
\begin{bmatrix}
\sigma_{yz}^k \\
\sigma_{xz}^k \\
\sigma_{zz}^k
\end{bmatrix}
=
\begin{bmatrix}
\overline{\sigma_{yz}} \\
\overline{\sigma_{xz}} \\
\overline{\sigma_{zz}}
\end{bmatrix}. \tag{S48}
$$

ii) Following the assumption of displacement continuity at layer interfaces, the in-plane strains are assumed uniform along the thickness, and thus equal to the in-plane strains of the overall material:

$$\begin{bmatrix} \epsilon_x^k \\ \epsilon_y^k \\ \epsilon_{xy}^k \end{bmatrix} = \begin{bmatrix} \overline{\epsilon_x} \\ \overline{\epsilon_y} \\ \overline{\epsilon_{xy}} \end{bmatrix}. \tag{S49}$$

iii) The layer thicknesses are small enough to assume that the stress and strain in every layer are uniform.

The averaging can thus be written as a simple weighted summation:

$$\frac{1}{V}\int_V A\,dV = \sum_{k=1}^N A^k v^k, \tag{S50}$$

Where $v^k = \frac{h^k}{H}$, $h^k$ is the thickness of a given layer and $H$ is the total thickness.

iv) The in-plane stresses are assumed to be averaged through the thickness of the overall material:

$$\begin{bmatrix} \overline{\sigma_x} \\ \overline{\sigma_y} \\ \overline{\sigma_{xy}} \end{bmatrix} = \sum_{k=1}^N v^k \begin{bmatrix} \sigma_x^k \\ \sigma_y^k \\ \sigma_{xy}^k \end{bmatrix}. \tag{S51}$$

v) The out-of-plane strains are assumed to be averaged through the thickness of the overall material:

$$\begin{bmatrix} \overline{\epsilon_{yz}} \\ \overline{\epsilon_{xz}} \\ \overline{\epsilon_{zz}} \end{bmatrix} = \sum_{k=1}^N v^k \begin{bmatrix} \epsilon_{yz}^k \\ \epsilon_{xz}^k \\ \epsilon_{zz}^k \end{bmatrix}. \tag{S52}$$

It is more convenient to reorganize the Voigt's relationship expressed in equation (S47) into separate in- and out-of-plane stresses and strains. It is achieved by using the transformation matrix:

$$P = \begin{bmatrix} 1 & 0 & 0 & 0 & 0 & 0 \\ 0 & 1 & 0 & 0 & 0 & 0 \\ 0 & 0 & 0 & 0 & 0 & 1 \\ 0 & 0 & 0 & 1 & 0 & 0 \\ 0 & 0 & 0 & 0 & 1 & 0 \\ 0 & 0 & 1 & 0 & 0 & 0 \end{bmatrix}, \tag{S53}$$

such that

$$P \begin{bmatrix} \sigma_x^k \\ \sigma_y^k \\ \sigma_z^k \\ \sigma_{xz}^k \\ \sigma_{yz}^k \\ \sigma_{xy}^k \end{bmatrix} = \begin{bmatrix} \sigma_x^k \\ \sigma_y^k \\ \sigma_{xy}^k \\ \sigma_{xz}^k \\ \sigma_{yz}^k \\ \sigma_z^k \end{bmatrix}, \qquad \text{and} \qquad P \begin{bmatrix} \epsilon_x^k \\ \epsilon_y^k \\ \epsilon_z^k \\ \epsilon_{xz}^k \\ \epsilon_{yz}^k \\ \epsilon_{xy}^k \end{bmatrix} = \begin{bmatrix} \epsilon_x^k \\ \epsilon_y^k \\ \epsilon_{xy}^k \\ \epsilon_{xz}^k \\ \epsilon_{yz}^k \\ \epsilon_z^k \end{bmatrix}. \tag{S54}$$

Applying this transformation on both sides of equation (S47):

$$P \begin{bmatrix} \sigma_x^k \\ \sigma_y^k \\ \sigma_z^k \\ \sigma_{xz}^k \\ \sigma_{yz}^k \\ \sigma_{xy}^k \end{bmatrix} = PC^k P^{-1} P \begin{bmatrix} \epsilon_x^k \\ \epsilon_y^k \\ \epsilon_z^k \\ \epsilon_{xz}^k \\ \epsilon_{yz}^k \\ \epsilon_{xy}^k \end{bmatrix}, \quad \text{and finally} \quad \begin{bmatrix} \sigma_x^k \\ \sigma_y^k \\ \sigma_{xy}^k \\ \sigma_{xz}^k \\ \sigma_{yz}^k \\ \sigma_z^k \end{bmatrix} = PC^k P^{-1} \begin{bmatrix} \epsilon_x^k \\ \epsilon_y^k \\ \epsilon_{xy}^k \\ \epsilon_{xz}^k \\ \epsilon_{yz}^k \\ \epsilon_z^k \end{bmatrix}. \tag{S55}$$

By using the index "I" for in-plane quantities and "S" for out-of-plane, or shear, quantities, equation (S55) can be written in the form:

$$\begin{bmatrix} \sigma_I^k \\ \sigma_S^k \end{bmatrix} = \begin{bmatrix} H_{II}^k & H_{IS}^k \\ H_{SI}^k & H_{SS}^k \end{bmatrix} \begin{bmatrix} \epsilon_I^k \\ \epsilon_S^k \end{bmatrix}. \tag{S56}$$

To determine the homogenized stiffness matrix, called $A$, that relates stress and strain in the overall material:

$$\begin{bmatrix} \overline{\sigma_I} \\ \overline{\sigma_S} \end{bmatrix} = \begin{bmatrix} \overline{A_{II}} & \overline{A_{IS}} \\ \overline{A_{SI}} & \overline{A_{SS}} \end{bmatrix} \begin{bmatrix} \overline{\epsilon_I} \\ \overline{\epsilon_S} \end{bmatrix}, \tag{S57}$$

Using the initial assumptions i) and ii), $\sigma_S^k = \overline{\sigma_S}$ and $\epsilon_I^k = \overline{\epsilon_I}$, and this equation becomes:

$$\begin{cases} \sigma_I^k = H_{II}^k \overline{\epsilon_I} + H_{IS}^k \epsilon_S^k & \text{(a)} \\ \overline{\sigma_S} = H_{SI}^k \overline{\epsilon_I} + H_{SS}^k \epsilon_S^k & \text{(b)} \end{cases}. \tag{S58}$$

Using equation (S58.b), $\epsilon_S^k$ can be expressed in terms of overall material stresses and strains $\overline{\sigma_S}$ and $\overline{\epsilon_I}$ and substituted into equation (S58.a):

$$\sigma_I^k = \left( H_{II}^k - H_{IS}^k H_{SS}^{k\,-1} H_{SI}^k \right) \overline{\epsilon_I} + H_{IS}^k H_{SS}^{k\,-1} \overline{\sigma_S}. \tag{S59}$$

Now by averaging over both sides, the shear strain experienced by the overall material can be computed:

$$\sum_{k=1}^{N} \nu^k \sigma_I^k = \overline{\sigma_I} = \left( \sum_{k=1}^{N} \nu^k \left( H_{II}^k - H_{IS}^k H_{SS}^{k\,-1} H_{SI}^k \right) \right) \overline{\epsilon_I} + \left( \sum_{k=1}^{N} \nu^k H_{IS}^k H_{SS}^{k\,-1} \right) \overline{\sigma_S}. \tag{S60}$$

The specific matrix components in (S57) can now be expressed as

$$\begin{cases} \overline{A_{SI}} = \left\{ \sum_{k=1}^{N} \nu^k \ H_{SS}^{k}{}^{-1} \right\}^{-1} \left\{ \sum_{k=1}^{N} \nu^k H_{SS}^{k}{}^{-1} H_{SI}^{k} \right\} \\[2em] \overline{A_{II}} = \sum_{k=1}^{N} \nu^k \left( H_{II}^{k} - H_{IS}^{k} H_{SS}^{k}{}^{-1} H_{SI}^{k} \right) + \left\{ \sum_{k=1}^{N} \nu^k H_{IS}^{k} H_{SS}^{k}{}^{-1} \right\} \left\{ \sum_{k=1}^{N} \nu^k \ H_{SS}^{k}{}^{-1} \right\}^{-1} \left\{ \sum_{k=1}^{N} \nu^k H_{SS}^{k}{}^{-1} H_{SI}^{k} \right\} \\[2em] \overline{A_{SS}} = \left\{ \sum_{k=1}^{N} \nu^k \ H_{SS}^{k}{}^{-1} \right\}^{-1} \\[2em] \overline{A_{IS}} = \left\{ \sum_{k=1}^{N} \nu^k H_{IS}^{k} H_{SS}^{k}{}^{-1} \right\} \left\{ \sum_{k=1}^{N} \nu^k \ H_{SS}^{k}{}^{-1} \right\}^{-1} \end{cases}$$

$$\text{(S61)}$$

By definition, the equivalent shear moduli of the material can be found using the relation $\bar{C} = P\bar{A}P^{-1}$. This can be done numerically. Sun [5] showed that, after cumbersome calculations, the effective out-of- and in-plane moduli are in fact given by:

$$G_{eff} = \left( \sum_n \frac{h_n/h}{G_n} \right)^{-1}, \tag{S62}$$

$$\mu_{eff} = \sum_n \mu_n \cdot \frac{h_n}{h}, \tag{S63}$$

where $h$ is the total thickness, $h_n$ is the thickness of the $n^{\text{th}}$ layer, $G_n$ is the out-of-plane modulus of the $n^{\text{th}}$ layer and $\mu_n$ is the in-plane modulus of the $n^{\text{th}}$ layer. In other words, the effective $G_{eff}$ is obtained from a weighted geometric average and the effective $\mu_{eff}$ from a weighted arithmetic average.